\documentclass[aps,pre,preprint]{revtex4-2} 

\usepackage{graphicx}
\usepackage{dcolumn}
\usepackage{bm}
\usepackage{amsmath,amssymb,amsfonts}
\usepackage[normalem]{ulem}
\usepackage{soul}
\usepackage[pdftex]{color}
\usepackage[font=footnotesize,labelfont=bf]{caption}
\usepackage{dcolumn}
\usepackage{xcolor}

\usepackage{comment}
\usepackage[caption=false]{subfig}
\usepackage{lipsum}

\usepackage{caption}
\begin{document}

\title{Spinodal decomposition in filled polymer blends exhibiting upper critical solution temperature behavior.}

\author{A.I. Chervanyov}

\affiliation{
 Institute of Theoretical Physics, University of M\"unster,
48149 M\"unster, Germany. \\ chervany@uni-muenster.de
}

\date{November 6, 2025}

\begin{abstract}
By extending the Sanchez–Lacombe lattice-fluid model for mixtures to the case of polymer blends containing solid fillers, we calculate the excess thermodynamic quantities arising from the presence of fillers. These results are then used to derive the spinodal stability condition of a filled polymer blend. In the low-compressibility limit, this condition reduces to a remarkably simple analytical expression that is derived self-consistently within the present framework. Comparison between the exact and approximate spinodal curves shows excellent agreement, with deviations in the spinodal temperature of less than $4~\mathrm{K}$, thereby validating the proposed approximation. The obtained analytical approximation enables a straightforward evaluation of the spinodal temperature without the extensive numerical calculations required to determine the exact spinodal condition. Both the exact and approximate spinodal conditions yield good quantitative agreement with experimental data for filled and unfilled blends.
\end{abstract}


\maketitle

\section{Introduction}
Despite its great practical significance, the effect of solid fillers on the stability and miscibility of polymer blends remains not fully understood, both experimentally and theoretically. Depending on the experimental conditions, the presence of fillers can either enhance the miscibility~\cite{NeterovLipatov,KitREW,KitMAC} of polymer blends or, conversely, reduce~\cite{LIPATOV_HORICHKO} their stability against phase separation.
The thermodynamic stability and miscibility of polymer blends are of central importance for their practical applications, as these characteristics govern the mechanical~\cite{Landry1991, PetiscoFerrero2017, Sangroniz2018, Paul2010}, optical~\cite{Takahashi2012, White2008, Paul2010}, and transport~\cite{Robeson2010, Jeong2023, Paul2010} properties of multicomponent polymer composites. Achieving controlled phase behavior in such systems is therefore a key step toward the development of advanced polymer composites with tailored functionalities. Understanding how the presence of fillers affect this behavior is therefore essential for both fundamental insight and practical design of polymer–particle based nanocomposites~\cite{Du2013,Chervanyov2023,Lee1999,Dorigato2024}.
In the present work, we investigate the relevant effects of fillers on the thermodynamic properties of binary polymer blends, with the aim of establishing a theoretical framework that enables the deliberate control of phase separation in polymer-particle composite materials.

Numerous theoretical and experimental studies have been devoted to modeling phase separation in polymer blends. Early approaches, such as the Flory–Huggins mean-field theory (FH)~\cite{Flory1953,Huggins1942}, provided the first quantitative framework for describing the entropy and enthalpy of mixing in polymeric systems. The FH theory, however, is not capable of predicting several important observable features of polymer blends, such as the lower critical solution temperature (LCST) behavior, without resorting to unphysical approximations for its main $\chi$ parameter. Nevertheless, FH has been used in most earlier theoretical studies~\cite{Lipatov1,Ginzburg}, discussed in detail in the next paragraphs, as the basis for describing filled polymer blends.

Later, more advanced theories incorporated compressibility and non-ideal volumetric effects through lattice-fluid~\cite{Sanchez1978,Sanchez1982} and other equation-of-state models~\cite{Lambert2000}. A pioneering role within this family of theories belongs to the Sanchez–Lacombe lattice-fluid theory~\cite{Sanchez1978,Sanchez1982}, which has been successfully applied to a wide range of polymer mixtures~\cite{Zoller1994,Radosz1985,Jackson1999}. Further progress in this field has led to the formulation of numerous equations of state (EoS) for polymer systems~\cite{Dee1988,Lambert2000}, as well as advanced theoretical treatments of nonathermal polymer blends~\cite{Lipson1993a,Lipson1993b,Freed1992,Freed1989}.

Despite the wide variety of models described above, which have provided significant progress in the field, only a few theoretical studies have so far addressed the effect of fillers on the thermodynamic state and stability of polymer blends. In his pioneering study, Warren~\cite{Warren} examined the influence of polymer polydispersity on phase separation in colloid–polymer mixtures. Conceptually, this work extended earlier analyses~\cite{LEKKERKERKER_POON_PUSEY} of monodisperse polymer–colloid systems based on the principles of free-volume theory. In that framework, the dense colloidal suspension provides a finite free volume accessible to polymer coils, which are modeled as noninteracting, spherical entities of fixed radius. These assumptions make Warren’s theory most applicable to dilute polymer solutions in dense colloidal media.
In contrast, the present work focuses on the opposite limit, i.e. systems containing a low concentration of solid fillers dispersed in dense, interacting polymer blends. Our model explicitly accounts for the excluded-volume interactions between the filler particles and the surrounding polymer matrix, which is crucial in determining the thermodynamic behavior of such systems.

A coarse-grained, pragmatically oriented, theoretical framework for describing filled polymer blends was first developed by Lipatov, Nesterov, and co-workers~\cite{Lipatov1,LIPATOV_HORICHKO,Lipatov2}. In their approach, FH for a polymer-polymer-solvent system was applied to rationalize the effect of fillers on the phase separation of this system. This simplified approach treats the fillers  as a continuum solvent-like phase interacting with the polymer components via the corresponding Flory–Huggins parameters. The translational entropy of polymer chains was neglected relative to the enthalpic interactions, which were assumed to dominate the demixing process. The main appeal of the Lipatov–Nesterov (LN) model lies in its conceptual simplicity and its qualitative ability to predict phase diagrams for polymer–polymer–particle systems. However, this model neglects several key physical aspects, including the finite size and shape of the fillers and specific polymer characteristics such as chain length or monomer volume. Consequently, the LN model cannot capture osmotic (excluded volume) or surface effects, both of which play a decisive role in determining the stability of polymer blends. As will be shown in the present work, the finite-size effect of the fillers neglected in the LN model plays a key role in explaining the effect of fillers on the miscibility of polymer blends.

A more advanced theoretical approach was later developed by Ginzburg~\cite{Ginzburg} and by He, Ginzburg, and Balazs~\cite{He}, who extended their and others earlier work in Ref.~[\onlinecite{Huh,ThompsonGinzburgMatsenBalazs}] on filled diblock copolymer systems to the case of polymer blends. The main advantage of this approach is that it explicitly accounts for the finite filler radius $R$ by introducing surface interaction terms in the free energy that scale as $R^2$. These terms were originally derived for diblock copolymer systems by analogy with the theory of filled polymer brushes~\cite{TangSolis}. In this theory, the entropic component of the surface interactions arises solely from the stretching of polymer chains near the interface~\cite{Huh,TangSolis}. While appropriate for brush-like structures formed in phase-separated block copolymer systems under strong-stretching conditions, this surface term is not directly applicable to polymer blends.
In dense polymer systems like realistic dry polymer blends, the polymer chains conform to nearly Gaussian statistics~\cite{Flory1974Lecture,Curro1987IntegralEquation}, which does not imply any preferred stretching direction. In addition, the model neglects the osmotic contribution $\sim R^3$ associated with the volume excluded to polymer chains by the fillers. As we demonstrate in the present work, this effect plays an important role in determining the miscibility of filled polymer blends. Furthermore, the model was formulated on the basis of the simplest version of the Flory–Huggins (FH) theory, which applies only to an idealized, strictly incompressible blend. In contrast, the present work employs the Sanchez–Lacombe equation of state (SL-EoS), which explicitly accounts for finite compressibility. The effect of compressibility is known~\cite{SLB,Rodgers,Voutsas} to be essential for an accurate description of the phase separation behavior in polymer blends.
 
The present work extends the SL-EoS to polymer blends containing solid, finite-sized, non-adsorbing~\cite{Chervanyov2013} fillers and establishes a consistent theoretical framework for describing the influence of these fillers on the thermodynamics and spinodal decomposition of these blends.
One essential advantage of this approach is that it incorporates the effects of finite compressibility, which are critical for accurately describing the thermodynamic behavior of real polymer blends. Furthermore, we derive a simple and practical analytical approximation for the spinodal in the low-compressibility limit of a polymer blend, which shows good agreement with both experimental data and the exact theoretical spinodal. This approximation allows a direct analytical evaluation of the spinodal curve, which otherwise requires full numerical minimization of the free energy~\cite{SLB}, a procedure that is computationally demanding even for unfilled polymer blends. Based on comparison with experiment, we demonstrate that the osmotic effect of fillers, incorporated here for the first time, can account for the experimentally observed compatibilizing effect of fillers in polymer blends exhibiting upper critical solution temperature (UCST) behavior.

The paper is organized as follows. Section~\ref{sec:Th} presents the theoretical formulation based on the Sanchez–Lacombe lattice-fluid framework and obtains expressions for the excess thermodynamic quantities induced by the presence of fillers, as well as the corresponding spinodal condition. Section~\ref{subsec:incom} presents the low-compressibility approximation and derives an analytical expression for the corresponding spinodal condition. Section~\ref{sec:res} compares the theoretical predictions with experimental data, demonstrating the predictive accuracy of the proposed model. Finally, Section~\ref{sec:conclusions} summarizes the main findings and outlines possible extensions of the theory.

\section{Theory}
\label{sec:Th}
\subsection{Stability of pure polymer blend}
\label{subsec:pure}
We consider a polymer blend composed of homopolymers of species $1$ ($2$), having polymerization degree $r_1$ ($r_2$). The blend is filled with $M_3$ spherical fillers of radius $R$ and volume $v_R \equiv 4\pi R^3/3$. The filler volume fraction $\varphi$ is assumed to be much smaller than that of the polymer components. Note that this assumption closely reflects typical experimental conditions~\cite{Hemmati2014,Dorigato2024,KitMAC,Lipatov1}, thereby justifying the use of the infinite dilution approximation with respect to the fillers. Under this approximation, the effective pair interactions (e.g., polymer-mediated interactions~\cite{PREsym,PREat,ChervanyovSoftMater2014}) between fillers are assumed to make negligible contributions to thermodynamic quantities. Consequently, the role of fillers in the phase transformations of the blend reduces to the effects associated with the immersion of a single filler in this blend, as described in detail in the current section.
As shown in our previous work~\cite{Polymers1,PRE20,JPolSci}, for weak interactions between fillers and polymers, the osmotic component of the filler immersion energy can be comparable to, or even dominate over, its enthalpic counterpart (e.g., adsorption energy of polymers onto fillers). In particular, the competition between the above osmotic and adsorption effects has been theoretically found~\cite{PRE20} to govern the localization of fillers in a diblock copolymer (DBC) system. Furthermore, the dominance of the osmotic effect, predicted~\cite{Chervanyov2025} to promote interfacial localization of fillers, provides a consistent explanation for the experimentally observed interfacial localization reported in Ref.[\onlinecite{KitMAC}]. 
	
Further theoretical calculations of the excess thermodynamic quantities caused by the presence of fillers rely on the Sanchez–Lacombe equation of state (SL-EoS) for a pure polymer blend. The main advantage of the SL-EoS over its Flory–Huggins counterpart (FH-EoS) is that the former considers a compressible lattice, thus allowing the description of a realistic compressible blend of polymers with different monomer sizes. As is well known~\cite{SLB}, it is the compressibility effects that underlie the lower critical solution temperature (LCST) behavior in polymer blends. In contrast to FH-EoS, SL-EoS can capture this effect without resorting to unphysical approximations for the FH $\chi$-parameter.
An additional advantage of the SL-EoS is that it explicitly incorporates the interaction energies of the pure polymer components (hereafter referred to as intra-interaction energies). These energies can be directly related to the parameters of the semi-empirical Tait equation~\cite{RodgersJAppPolySci1993}, thereby providing important input experimental parameters for the theory~\cite{Lipson1}. Recall that the cumulative effect of enthalpic interactions in the FH theory, in contrast,  is described by a single FH $\chi$-parameter, the approximation justified only near the athermal limit. As will be shown below, the advantages of the SL-EoS described above are crucial for an adequate description of the effects of fillers on the phase separation behavior of polymer blends.

The SL-EoS~\cite{SL,SLB} can be conveniently represented in the form
\begin{equation}
\label{SL_P}
\tilde{P}_0 = \eta \left( p + \eta Q(\eta) \right) - \frac{1}{2} \beta \eta^2 z \epsilon(\phi),
\end{equation}
where $\tilde{P}_0 = v^* \beta P_0$ is the reduced pressure, $P_0$ being the pressure of a pure polymer blend, $v^{*-1} = \sum_{i=1,2} \phi_i / v_i$ is the harmonic average of the monomer volumes $v_1$ and $v_2$, and $\beta = (kT)^{-1}$. Here, $k$ and $T$ are the Boltzmann constant and temperature, respectively, while $\phi_1 \equiv \phi$ and $\phi_2 \equiv 1 - \phi$ denote the volume fractions of the polymer blend components.
Further, $z$ is the lattice coordination number; $\eta = \eta_1 + \eta_2$, where $\eta_i$ is the volume fraction of monomer $i$;
$p = \sum_{l=1,2} \phi_l r_l^{-1}$,
$Q(\eta) = -(\eta + \log(1 - \eta)) / \eta^2$, and
$\epsilon(\phi) = \sum_{m,l=1,2} \epsilon_{ml} \phi_m \phi_l$  is the total polymer interaction energy, $\epsilon_{ml}$ quantifies the cohesive energy between polymer species $m$ and $l$ ($m,l=1,2$).

The chemical potential $\mu_i^{0}$ ($i=1,2$) of the polymer molecules of species $i$ can be readily derived from Eq.~(\ref{SL_P}), yielding
\begin{gather}
\label{SL_mu}
\beta \mu_i^{0} =
\left[
\log(\eta \phi_i) + 1 - p r_i + r_i (\eta^{-1} - 1) \log(1 - \eta)
\right] \\ \nonumber
+ r_i \eta \left(
\eta^{-2} \tilde{P}_0 - (1 - \phi_i) (\chi(1 - \phi_i)  + \beta z \epsilon_{ii}/2)
\right),
\end{gather}	
where 	
$
\chi = z\beta \left(2 \epsilon_{12} - \epsilon_{11} - \epsilon_{22}  \right)/2
$
is the FH interaction parameter~\cite{Rubinstein2003}. 

The standard stability condition for a polymer blend~\cite{Gugen} is expressed as $\partial_{M_1} \mu_1^{0}(P, M_1, M_2) > 0$, where  $M_1$ ($M_2$) is the number of polymers of species $1$ ($2$). This condition can, in the general case, be conveniently recast as
\begin{equation}
\label{spin0}
S_0\equiv\partial_{\phi} \left( r_1^{-1}\mu_1 - r_2^{-1}\mu_2 \right)
-
\eta^{-1} \tilde{\kappa}_0 \left( \partial_{\phi} \tilde{P}_0 \right)^2 \ge 0,
\end{equation}
as derived in Ref.~[\onlinecite{Chervanyov2023}]. Here, 
\begin{equation}
\label{comp0}
\tilde{\kappa}_0 \equiv
\eta^{-1}\left(
\eta(1 - \eta)^{-1} + p -\beta z \eta \epsilon
\right)^{-1}
\end{equation}
is the reduced isothermal compressibility of a pure polymer blend. The advantage of the representation given by Eq.~(\ref{spin0}) is that the effect of the finite compressibility of the blend is entirely described by the second term on the left-hand side (l.h.s.) of this inequality. In the incompressible limit $ \tilde{\kappa}_0\to 0$, this term vanishes.

For the considered case of SL-EoS, the equation $S_0 = 0$, which determines the spinodal curve delineating the stability and instability regions in the $\phi$–$T$ plane, can be readily obtained from the general expression for $S_0$ given in Eq.~(\ref{spin0}). A straightforward substitution of Eqs.~(\ref{SL_P}) and~(\ref{SL_mu}) into Eq.~(\ref{spin0}) transforms the spinodal equation into the form
\begin{equation}
\label{spin00}
\sum\limits_{l=1}^{2}(r_l\phi_l)^{-1} + 2\eta\chi = \eta^{-1} \tilde{\kappa}_0 \Pi_{0}^2,
\end{equation}
where $\Pi_{0} = \eta  \left(r_1^{-1} - r_2^{-1} - \beta \eta \epsilon'(\phi)\right)$ and
the prime denotes differentiation with respect to the argument.

Note that the spinodal equation given by Eq.~(\ref{spin00}) is not sufficient to determine the spinodal curve $T(\phi)$. This equation must be complemented by  SL-EoS, which allows $\eta$ to be expressed as a function of $\phi$ and $T$. The additional relationship among $\eta$, $\phi$, and $T$ is obtained by numerically solving  SL-EoS, written in the form
\begin{equation}
\label{Eq2_0}
 Q(\eta) - \eta^{-2}\tilde{P}_0 = \tfrac{1}{2}\beta z \epsilon(\phi) - \eta^{-1}p,
\end{equation}
with respect to $\eta$. The reduced pressure $\tilde{P}_0$ in Eq.~(\ref{SL_P}) is treated as a fixed quantity corresponding to the chosen experimental conditions. At room temperature and atmospheric pressure, $\tilde{P}_0$ evaluates to $3.06 \times 10^{-3}$.
Hereafter, we set the coordination number $z=6$, which corresponds to the simple cubic lattice. 

The spinodal curve can be readily evaluated from simultaneous Eqs.~(\ref{Eq2_0}) and~(\ref{spin00}) for any given set of experimental parameters $r_i$ and $\epsilon_{ij}$ ($i, j = 1, 2$). The cross-interaction parameter $\epsilon_{12}$, which characterizes the interactions between dissimilar polymer species, is commonly estimated using the Berthelot combining rule, $\epsilon_{12} = \sqrt{\epsilon_{11}\epsilon_{22}}$. In the present work, however, we adopt a temperature-dependent form,
\begin{equation}
\label{e12}
\epsilon_{12} = \epsilon_{12}^0 + a kT,
\end{equation}
which corresponds to the widely used~\cite{Rubinstein2003}
 approximation for the Flory–Huggins 
$\chi$-parameter,
$\chi=za/2 + b/T$, where $a$ and $b$ represent the entropic and enthalpic contributions, respectively. 
\subsection{Contribution of fillers to the thermodynamics and stability of a polymer blend}
\label{subsec:fil}
Immersion of fillers into the polymer blend introduces corrections to the thermodynamic quantities given by Eqs.~(\ref{SL_P}) and~(\ref{SL_mu}). These corrections, in turn, modify the SL-EoS defined by Eq.~(\ref{SL_P}). To derive the equation of state that accounts for the presence of fillers, we assume semi-grand canonical conditions, i.e., constant chemical potentials of the blend components at fixed volume and temperature, with a constant number of fillers $M_3$. Under these conditions, the correction $\Delta \tilde{P}$ to the reduced pressure of the blend due to the purely osmotic effect of fillers is given by~\cite{Chervanyov2023}
\begin{equation}
		\label{Pfil}
\Delta \tilde{P} =
		\varphi\eta (1-u + \tilde{\kappa}_0^{-1}),
	\end{equation} 
which brings the equation of state into the form
\begin{equation}	
		\label{EoS}
 \tilde{P} = \eta\left[
 \eta Q(\eta) +  \varphi (1-u + \eta^2 (1 - \eta)^{-1})
 + p(1+\varphi \eta ) - \frac{z}{2} \beta \eta  \epsilon(\phi) (1+2\varphi \eta )		
\right].
	\end{equation} 
	
Here, $\tilde{P} = v \beta P$ is the reduced pressure, $\varphi = M_3v_R/Mrv$ the hard core volume fraction of fillers, $Mr=\sum_{i=1,2} M_ir_i$, and
$u = 1-v/v_R$ is the parameter quantifying the asymmetry in volumes of fillers and monomers. Note that for typical experimental conditions $u\sim 1$. 
Hereafter, we assume the same monomer hard core volumes $v_1=v_2=v$. Note that the first term proportional to $1-u$ in the r.h.s. of Equation~(\ref{Pfil}) arises from the translational entropy of fillers. 
	
The chemical potential of fillers, straightforwardly derived from Eq.~(\ref{Pfil}), is given by 	
	\begin{equation}
		\label{mu3}
		\beta\mu_3 = \log(\phi\eta) + \tilde{P}_0(\eta, \phi)/(1-u).
	\end{equation} 
According to Eq.~(\ref{mu3}), the chemical potential of fillers consists of two contributions. The first contribution, represented by the first term on the right-hand side (r.h.s.) of  Eq.~(\ref{mu3}), originates from the translational entropy of the fillers. The second term describes the minimal work required to create a cavity of volume $v_R$ against the pressure $P_0$ exerted by the polymer blend.

By integrating the Maxwell relations
$\partial_{M_3}\mu_i = \partial_{M_i}\mu_3$,
the excess chemical potential $\Delta \mu_i$ of the polymer blend component $i$ arising from the presence of fillers can be readily obtained. In the limit of infinite dilution with respect to the fillers, one finds
\begin{equation}
\label{mu1}
\beta \Delta \mu_1 = \eta \left( 1 + r_1 \eta (1 - \eta)^{-1}
- \beta r_1 \eta (2 \epsilon + \phi_2 \epsilon'(\phi_1)) \right)
\end{equation}
$\Delta \mu_2$ can be deduced from Eq.~(\ref{mu1}) by a straightforward interchange of the indices. This equation describes the perturbation of the thermodynamic state of the pure blend induced by the coupling of the fillers to its local pressure field $\tilde{P}_0(\eta, \phi)$. The correction term $\Delta \mu_1$ quantifies this perturbation by linking the structural parameters ($\eta$ and $\phi$) to the thermodynamic response of the filled blend.

In the considered infinite-dilution approximation, the presence of filler particles modifies the simple stability condition given by Eq.~(\ref{spin0}) by introducing a correction proportional to the filler fraction $\varphi$. To derive this correction, we use the standard stability condition for a three-component system~\cite{Gugen} 
\begin{equation}
\label{spin_aux}
\partial_{M_1} \mu_1(\{M_i\}, P)\partial_{M_3} \mu_3(\{M_i\}, P)
- \partial_{M_1} \mu_3(\{M_i\}, P)\partial_{M_3} \mu_1(\{M_i\}, P) \geq 0
\end{equation}
there $\{M_i\}$ denotes a set of number of molecules of all species and $P\equiv P_0 + \Delta P$ is the pressure of filled polymer system. It proves convenient to recast the above spinodal condition in terms of the reduced derivatives $m_{ij} = M_j \partial_{M_{j}} \mu_i(\{M_i\},P)$:
\begin{equation}
\label{spin}
S \equiv m_{11} - \left( m_{31}/m_{33} \right) m_{13} \geq 0.
\end{equation}
In this form, the correction to the stability condition of the pure polymer blend, given by Eq.~(\ref{spin0}), caused by the presence of fillers, is given by the second term on the r.h.s. of Eq.~(\ref{spin}). It should also be noted that an additional contribution to this correction originates from the first term, since $m_{11}$ differs for the filled and pure blends.

To rationalize the spinodal condition given by Eq.~(\ref{spin}), one needs explicit expressions for the reduced derivatives $m_{ij}$ of the chemical potentials at constant pressure. Changing the independent variables from $(P,\{M_i\})$ to $(\eta,\phi,\varphi)$ and differentiating yields rather cumbersome expressions, which are derived in Appendix~\ref{app:A}. Substituting these results into the left–hand side (l.h.s.) of Eq.~(\ref{spin}) gives the explicit spinodal equation $S(\eta,\phi,\varphi)=0$.

Recall that, in addition to the independent variables $(\eta, \phi, \varphi)$, the function $S$ depends on five material parameters, $r_i$ and $\epsilon_{ij}$ ($i,j = 1,2$). While four of these parameters, $r_i$ and $\epsilon_{ii}$ ($i = 1, 2$), are typically known for most polymer blends, the cross-interaction parameter $\epsilon_{12}$ is assumed to have the form given by Eq.~(\ref{e12}). The parameters $\epsilon_{12}^0$ and $a$ defining $\epsilon_{12}$ can be determined from the experimental spinodal of a pure polymer blend, as described in Sec.~\ref{sec:res}.

The spinodal condition given by Eq.~(\ref{spin}) is not sufficient to determine the spinodal curve in the conventional form $T(\phi)$ for a given $\varphi$, because, in addition to $T$ and $\phi$, the function $S$ in this condition also depends on a third variable, $\eta$. An additional relationship among $T$, $\phi$, and $\eta$ can be obtained from the equation of state given by Eq.~(\ref{EoS}), which can be conveniently recast in the form
\begin{equation}
\label{Eq2_1}
Q(\eta) - \eta^{-2}\tilde{P} + \varphi  (\eta^{-1}(1-u) + \eta (1-\eta)^{-1}) =
\beta z \epsilon(\phi) (2^{-1} + \varphi\eta) - p \eta^{-1}(1 + \varphi\eta).
\end{equation}
Recall that the reduced pressure $\tilde{P}$ in the l.h.s of Eq.~(\ref{Eq2_1}) must be treated as a fixed quantity determined by given experimental conditions. In the present calculations, we set $\tilde{P} = 3.06 \times 10^{-3}$, corresponding to room temperature and atmospheric pressure.

The spinodal curve can be obtained by simultaneously solving Eq.~(\ref{Eq2_1}) and the spinodal condition $S = 0$, with $S$ defined in Eq.~(\ref{spin}), for any given set of experimental input parameters $r_i$ and $\epsilon_{ii}$ ($i = 1, 2$), and the cross-interaction parameter $\epsilon_{12}$ [Eq.~(\ref{e12})], determined by the procedure described in Sec.~\ref{sec:res}.
\subsection{Low-compressibility limit of a polymer blend}
\label{subsec:incom}

As will be discussed in detail in Sec.~\ref{sec:res}, the upper critical solution temperature (UCST) behavior of polymer blends occurs, in particular, when the blend is only weakly compressible. It is therefore instructive to simplify the exact spinodal condition, given by the simultaneous Eqs.~(\ref{spin}) and~(\ref{Eq2_1}), by employing the above low-compressibility assumption. In the leading order with respect to the compressibility, one arrives at the limit of an incompressible polymer blend, which corresponds to the condition of vanishing isothermal compressibility, $\tilde{\kappa}_0 = 0$. Within the framework of the present model, this condition can be obtained in explicit form from the expression for the compressibility given by Eq.~(\ref{comp0}) by setting it to zero, which immediately yields $\eta \to 1$. This result is physically natural, since $\eta$ represents the fraction of the total volume occupied by polymer segments; hence, the limit $\eta \to 1$ corresponds to the situation in which no free volume remains and any further compression becomes impossible.

In what follows, we obtain the low-compressibility limit (hereafter referred to as incompressible limit) of the spinodal for a filled polymer blend by carefully taking the limit $\eta \to 1$ in the exact spinodal condition, Eq.~(\ref{spin}), rather than by imposing the incompressibility constraint directly on the free energy, as done in Ref.~[\onlinecite{Ginzburg}]. This approach allows us to correctly account for the osmotic effect of fillers on the spinodal.

Taking the limit $\eta \to 1$ brings the spinodal condition given by Eq.~(\ref{spin}) to remarkably simple analytical form
\begin{equation}
\label{spin_inc}
S_{inc}= \sum_{l=1,2}  (\phi_l r_l)^{-1} + 2\chi  + 
\varphi \left( \Pi_{inc}'(\phi) + \Pi_{inc}(\phi)^2  \right)  > 0,
\end{equation}
where $\Pi_{inc} \equiv \partial_{\phi} \tilde{P}_0 (\eta=1) = r_1^{-1} - r_2^{-1} - \beta\epsilon'(\phi) $ is the incompressible limit of the derivative of  the reduced blend pressure with respect to the fraction $\phi$ quantifying the composition of the blend. 

Setting $\varphi = 0$ reduces the condition given by Eq.~(\ref{spin_inc}) to the well-known spinodal condition~\cite{SL} for an incompressible pure polymer blend. The presence of fillers introduces corrections to this condition, caused by the dependence of the filler chemical potential on the blend pressure, as can be elucidated from Eq.~(\ref{mu3}). Note that this correction depends only on the derivatives of the blend pressure with respect to $\phi$, which describe how the pressure of the polymer matrix responds to variations in its composition. Remarkably, even in the incompressibility limit $\eta \to 1$, these pressure derivatives remain finite and can be straightforwardly evaluated from the SL-EoS given by Eq.~(\ref{SL_P}).

Physically, the above correction introduced by the fillers represents the mechanical work required to maintain the same number of hard, incompressible inclusions when the local pressure of the blend changes due to composition fluctuations. Because each filler occupies a fixed excluded volume $v_R$, such fluctuations force the surrounding polymer matrix to perform additional work of order $P_0 v_R$, thereby increasing the energetic cost of composition variations. This mechanical stiffening of the system manifests as the correction term in the spinodal condition proportional to $\partial_{\phi}^2 P_0$, which describes the additional stabilization of the homogeneous phase imposed by the fillers. The term in Eq.~(\ref{spin_inc}), proportional to $(\partial_{\phi} P_0)^2$, has a more complex dependence on the blend composition and can therefore either promote or reduce the stability of the mixture. The relative magnitudes of these two terms and their combined effect on the spinodal depend on the experimental parameters $r_i$ and $\epsilon_{ii}$ ($i = 1, 2$), as well as on the temperature dependence of the cross-interaction parameter $\epsilon_{12}$. 

The analytical approximation for the spinodal of a filled UCST polymer blend, given by Eq.~(\ref{spin_inc}), demonstrates that the filler-induced correction originates from the dependence of the immersion energy on blend composition, reflecting the mechanical work required for the matrix to act against the filler excluded volume. Another important observation is that this correction depends not only on the $\chi$ parameter, which quantifies the difference between cross-species and intra-species cohesive energies, but also on the difference $\epsilon_{11}-\epsilon_{22}$, which characterizes the energetic asymmetry of the pure polymer components forming the blend.

\section{Comparison with experiment}
\label{sec:res}
The spinodal condition derived in Sec.~\ref{subsec:fil} requires important experimental input in the form of the parameters $r_i$ and $\epsilon_{ii}$ ($i = 1, 2$). In the present work, we focus on investigating the effect of fillers on the upper critical solution temperature (UCST) behavior of polymer blends. Accordingly, as the necessary experimental input, we use the set of parameters corresponding to the polymers used in Ref.~\onlinecite{Hemmati2014} to investigate this behavior.

In the experimental system, studied in  Ref.~\onlinecite{Hemmati2014}, high density polyethylene (HDPE) and ethylene–vinyl acetate (EVA) blends were combined with organically modified nanoclay to examine the influence of nanoparticles on blend miscibility, interfacial tension, and phase separation kinetics. The objective of the study was to understand how the incorporation of organoclay alters the thermodynamic and kinetic behavior of polymer blends exhibiting the UCST behavior. Specifically, the experiments demonstrated that introducing intercalated nanoclay lowers the phase separation temperature, modifies the morphology of the coexisting phases, and significantly affects the rate of phase separation due to restricted polymer chain diffusion in the presence of nanoparticles. The nanoclay (NC) loading was fixed at 3~wt\%, which corresponds to a volume fraction of approximately $\phi_{\mathrm{NC}} \simeq 0.012$, given the densities of EVA, HDPE, and NC as $0.94$, $0.95$, and $1.9~\mathrm{g/cm^3}$, respectively.

Although the EVA used in the experiments (hereafter referred to as EVA28) is a copolymer containing 28~wt\% vinyl acetate, it can nevertheless be represented within the theoretical model as a homopolymer characterized by a single inter-monomer interaction parameter. This approximation is justified because the vinyl acetate segments in EVA28 are randomly distributed along the chain, preventing the formation of distinct polar and nonpolar blocks and thus eliminating any significant intrachain segregation. Consequently, the interactions between ethylene and vinyl acetate segments are effectively averaged, making them weaker than the dominant intermolecular interactions between EVA28 and HDPE segments. Accordingly, within the framework of the SL-EoS, the inter-monomer interaction in EVA28 is used to be described~\cite{RodgersJAppPolySci1993} by a single parameter $\epsilon = k \times 685~\mathrm{K}$, whereas for HDPE it is given by $\epsilon = k \times 736~\mathrm{K}$.

To determine the spinodal $T(\phi)$, the simultaneous Eqs.~(\ref{spin}) and~(\ref{Eq2_1}) must be solved numerically for the selected set of experimental parameters, as explained in Sec.~\ref{subsec:fil}.  Recall that these equations include also the cross-interaction parameter $\epsilon_{12}$, which has a prescribed temperature dependence given by Eq.~(\ref{e12}) and contains two adjustable constants, $\epsilon_{12}^0$ and $a$. To determine these constants for the given experimental system, we fitted the spinodal curve $T(\phi)$, theoretically obtained by simultaneously solving Eqs.~(\ref{spin0}) and~(\ref{Eq2_0})
to the experimental data points reported in Ref.~\onlinecite{Hemmati2014} for the case of unfilled polymer blend. In the present and all subsequent calculations, the coordination number was set to $z = 6$, corresponding to a simple cubic lattice.

\begin{figure}[htbp]
\centering
\includegraphics[width=0.75\linewidth]{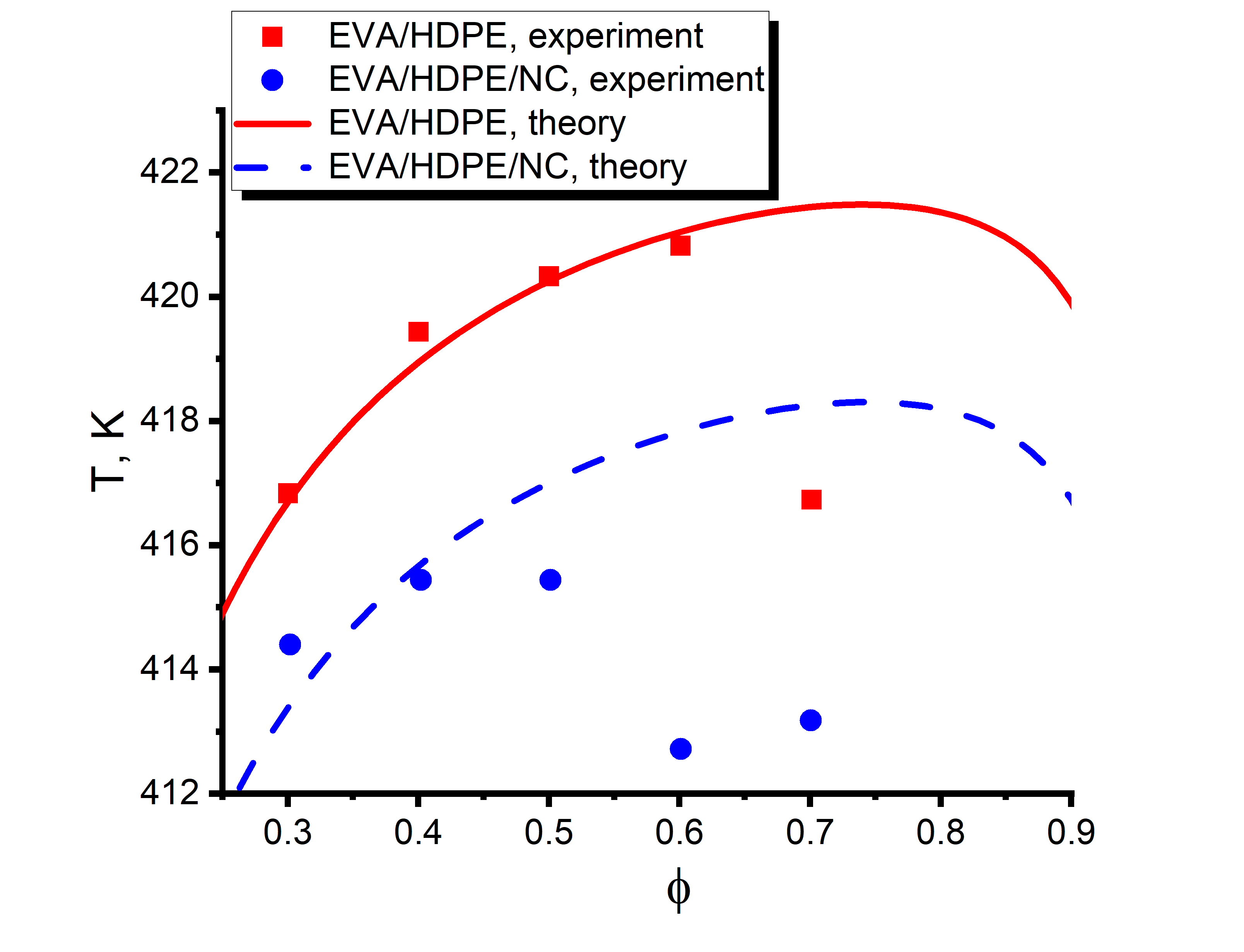}
\caption{
Comparison between experimental and theoretical spinodals of unfilled (EVA/HDPE) and filled (EVA/HDPE/NC) polymer blends. 
Symbols represent experimental data~\cite{Hemmati2014}, while solid and dashed lines correspond to the theoretical predictions obtained from Eqs.~(\ref{spin}) and~(\ref{Eq2_1}) for unfilled ($\varphi=0$) and filled blends, respectively. 
The presence of nanoclay (NC) lowers the spinodal temperature by approximately $5~\mathrm{K}$, in quantitative agreement with experimental observations. 
}
\label{fig1}
\end{figure}

The results of the described fitting procedure are shown in Fig.~\ref{fig1}. As can be seen from this figure, the temperature dependence of $\epsilon_{12}$ given by Eq.~(\ref{e12}), assumed on physically plausible grounds, yields good agreement between the experimental results and theoretical predictions. The values of  parameters $\epsilon_{12}^0/k$ and $a$ that provide the best fit are $707.151~\mathrm{K}$ and $0.00785$, respectively. Remarkably, the value of $\epsilon_{12}/k = 710.0 K$, obtained by substituting these parameters into Eq.~(\ref{e12}), agrees very well with the Berthelot combining rule $\epsilon_{12}/k = \sqrt{\epsilon_{11}\epsilon_{22}}/k = 709.5~\mathrm{K}$, which is derived from the interaction parameters of the pure components. The Berthelot approximation is therefore well justified in the case where $\epsilon_{12}$ is assumed to be temperature independent.

Using the obtained approximation for $\epsilon_{12}$, one can compare the theoretical predictions for the effect of fillers on the spinodal of polymer blends with the experimental results. The outcome of this comparison is presented in Fig.~\ref{fig1}. As can be seen from this figure, both experiment and theory predict a reduction in the spinodal temperature due to the presence of fillers, with a maximum decrease of approximately $\sim 7~\mathrm{K}$. The theoretical predictions show good agreement with the experimental data for smaller volume fractions, $\phi_1 < 0.5$, while a larger deviation $\sim 4~\mathrm{K} $ is observed at higher $\phi_1$ values. It can therefore be concluded that the developed theory provides an adequate explanation of the experimental observations without the need for any adjustable parameters, apart from $\epsilon_{12}$, which was determined independently by fitting the spinodal of the pure (unfilled) polymer system as described in the preceding paragraph.

\begin{figure}[htbp]
\centering
\includegraphics[width=0.75\linewidth]{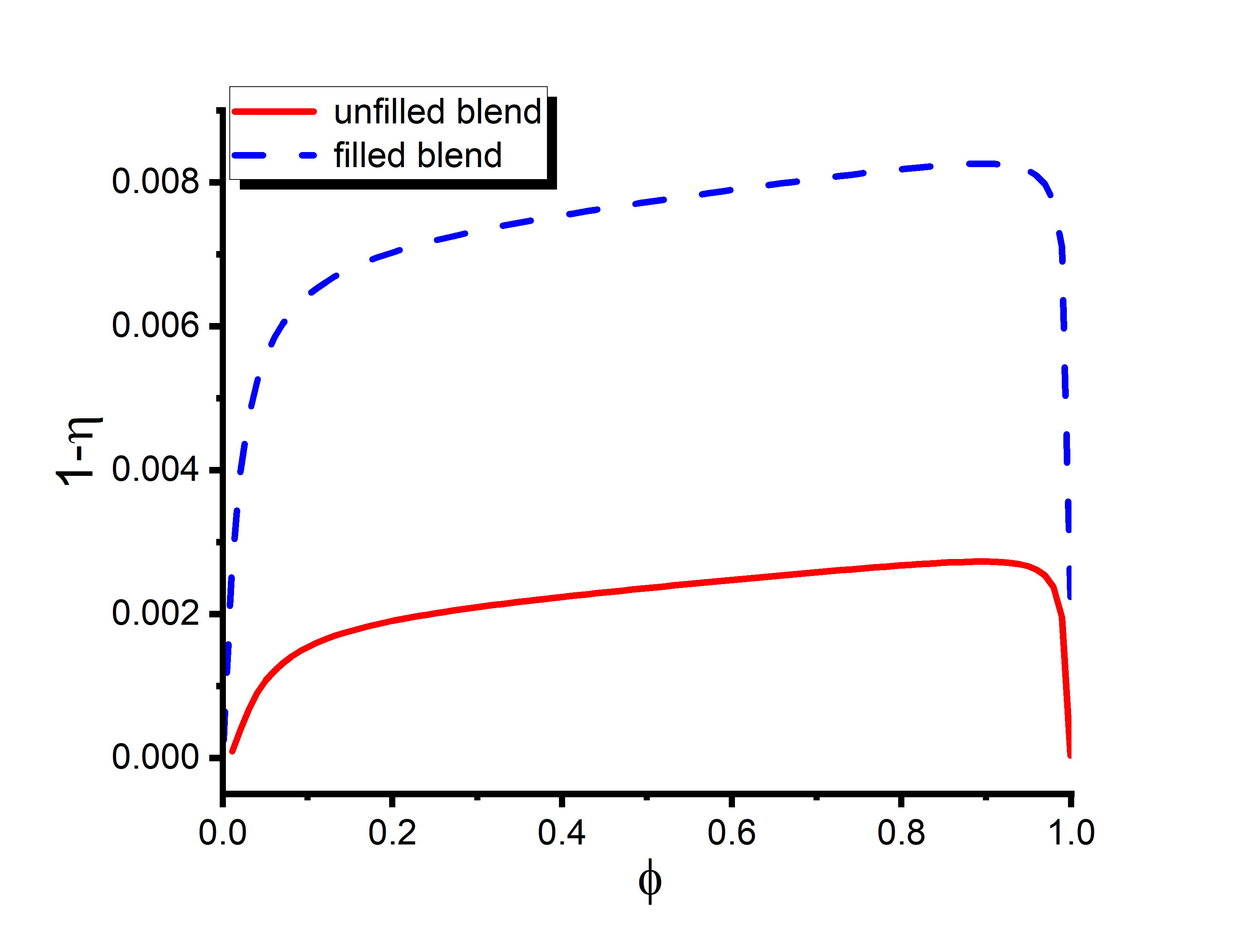}
\caption{
Dependence of the free volume, defined as $1 - \eta$, on the EVA fraction $\phi$ at the spinodal for unfilled and filled polymer blends. 
The solid and dashed lines correspond to the theoretical predictions for the unfilled and filled blends, respectively. 
}
\label{fig2}
\end{figure}

The good agreement obtained between the experimental spinodal data and theoretical predictions confirms the reliability of the developed model. It is therefore instructive to extend these predictions to other characteristics of the blend in the presence of fillers. Fig.~\ref{fig2} shows the dependence of the free volume, defined as 
$1-\eta$, on the EVA fraction $\phi$ at the spinodal. This quantity represents the relative fraction of excess volume available to the polymers but not occupied by them. As seen in the figure, within the experimentally accessible range of EVA fractions, $0.03 < \phi < 0.97$, the free volume increases slightly and nearly linearly with $\phi$, reaching a maximum value of $0.0083$ for the filled blend and $0.0025$ for the pure blend. Interestingly, the free volume is slightly larger in the filled polymer blend compared to the pure one. In both cases, however, these values remain small, not exceeding $0.9~\%$ of the total volume available to the polymers. This small free volume underlies the UCST behavior of the system, as it limits the effects associated with the finite compressibility of the blend.

\begin{figure}[htbp]
\centering
\includegraphics[width=0.75\linewidth]{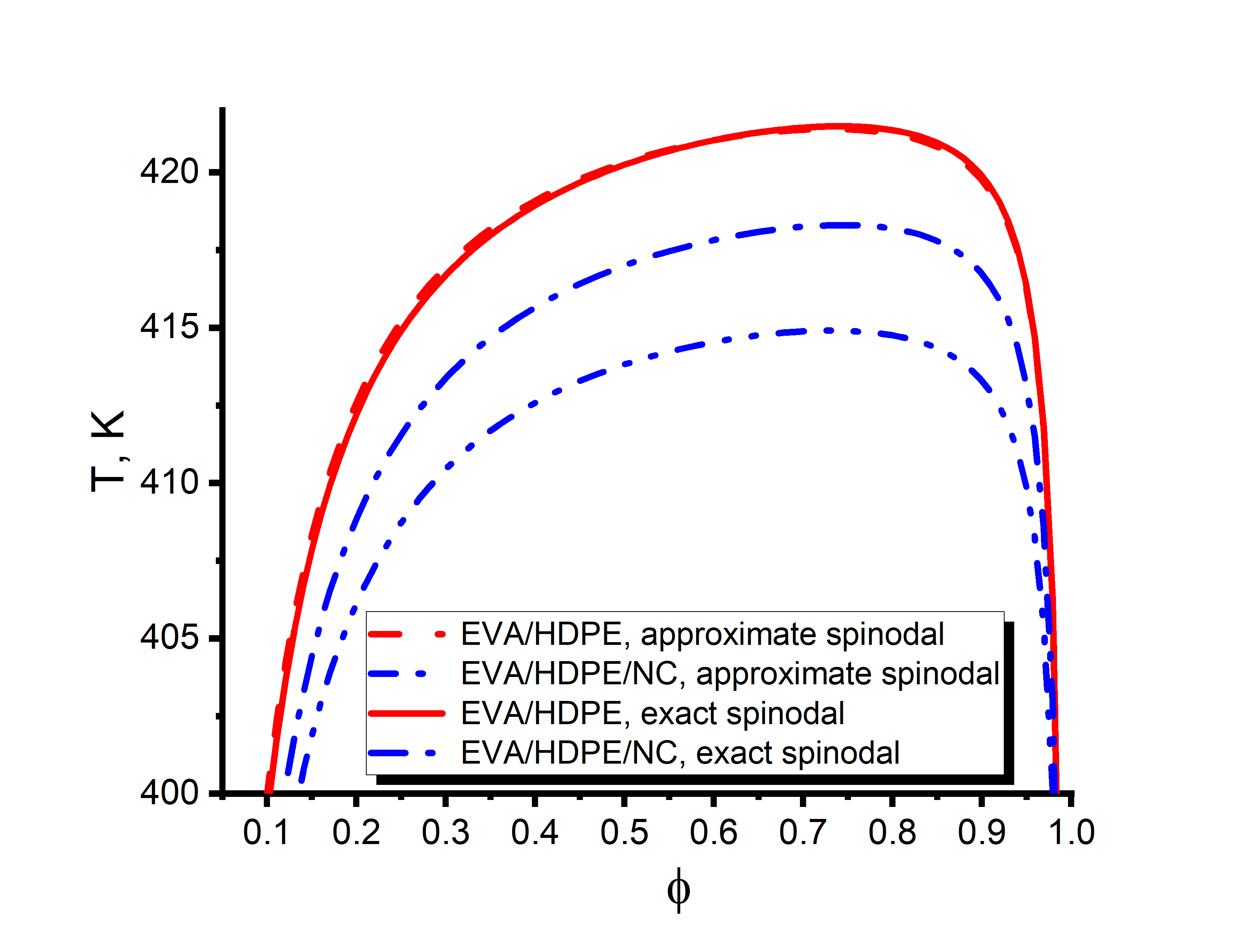}
\caption{
Comparison between the exact and approximate spinodals for unfilled (EVA/HDPE) and filled (EVA/HDPE/NC) polymer blends. 
The exact spinodals (solid and dash-doted lines) are obtained from the full numerical solution of Eqs.~(\ref{spin}) and~(\ref{Eq2_1}), whereas the approximate spinodals (dashed and dash-dot-doted lines) are calculated from the analytical incompressible-limit expression, Eq.~(\ref{spin_inc}). 
}
\label{fig3}
\end{figure}

As the compressibility effects are suppressed under the considered experimental conditions, it is justified to apply the approximation of a totally incompressible polymer blend developed in Sec.~\ref{subsec:incom}. Within this approximation, the spinodal temperature $T(\phi)$ can be evaluated analytically using the derived Eq.~(\ref{spin_inc}). The comparison between the exact spinodal, obtained through the detailed procedure described at the beginning of this section, and that predicted by the simplified Eq.~(\ref{spin_inc}) is shown in Fig.~\ref{fig3}. This figure demonstrates that the spinodal obtained using the incompressible approximation shows remarkably good agreement with its exact counterpart for the unfilled polymer blend. In the case of the filled blend, the incompressibility approximation slightly underestimates the spinodal temperature, with a maximum deviation of about $4~\mathrm{K}$. Overall, the incompressible-limit spinodal provides an excellent approximation to the exact result under the UCST experimental conditions, confirming the validity of the incompressibility approximation for these systems.

\begin{figure}[htbp]
\includegraphics[width=0.75\linewidth]{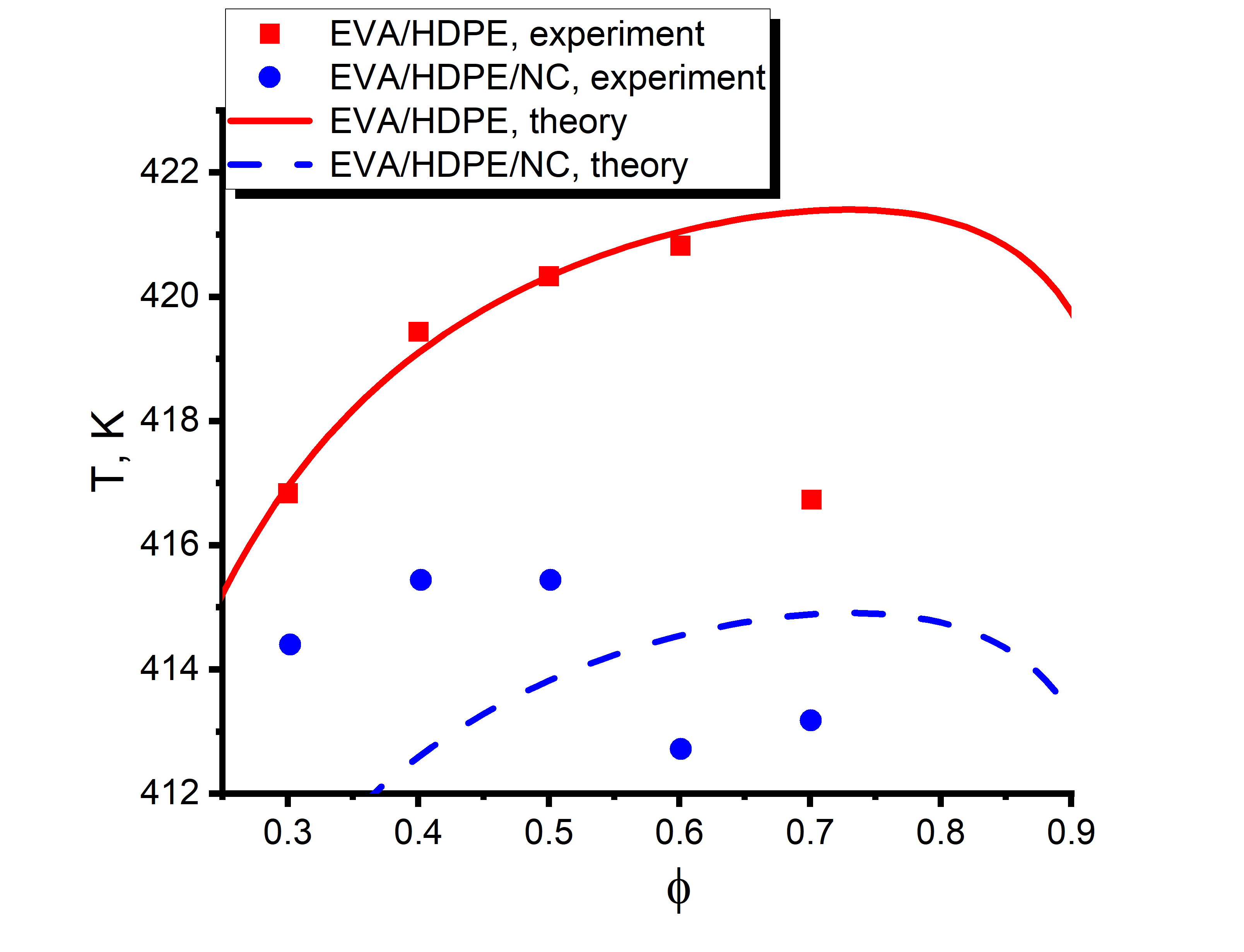}
\caption{
Comparison between the experimental spinodals and the approximate theoretical spinodals  obtained from Eq.~(\ref{spin_inc}) for unfilled (EVA/HDPE) and filled (EVA/HDPE/NC) polymer blends. 
Symbols represent the experimental data~\cite{Hemmati2014}, while the solid and dashed lines correspond to the approximate theoretical spinodals for the unfilled and filled blends, respectively. 
}
\label{fig4}
\end{figure}

Fig.~\ref{fig4} compares the approximate theoretical spinodal, given by Eq.~(\ref{spin_inc}), with the same experimental data shown in Fig.~\ref{fig1}. The adjustable constant (effective cross-interaction temperature) $\epsilon_{12}^0/k$ was slightly increased by $0.033~\mathrm{K}$ to achieve the best fit to the experimental data, resulting in $\epsilon_{12}^0/k = 707.151~\mathrm{K}$. As seen in Fig.~\ref{fig4}, the approximate spinodal obtained for the filled polymer blend provides good agreement with the experimental data across the entire range of $\phi$. Similar to its exact counterpart, the approximate spinodal reproduces the experimental trend very well for $\phi < 0.6$, while at higher $\phi$ the experimental $T(\phi)$ decreases more steeply than predicted. Overall, the approximate spinodal calculated in the incompressible limit shows good agreement with experiment and can therefore be used as a reliable, computationally efficient alternative to the exact spinodal in practical evaluations.

\section{Conclusions}
\label{sec:conclusions}
In the present work, a theoretical framework is developed to describe the thermodynamic behavior of polymer blends containing solid fillers by extending the Sanchez–Lacombe lattice-fluid model~\cite{SL} for mixtures to the case of filled polymer blends. The theoretical analysis focuses on the influence of filler incorporation on the phase stability and spinodal decomposition of polymer blends exhibiting upper critical solution temperature (UCST) behavior. The model accounts for the changes in the relevant thermodynamic quantities induced by the fillers and establishes both analytical and numerical procedures for determining the spinodal temperature of filled polymer blends.

Relying on the expressions for the pressure and chemical potential of a pure polymer blend respectively given by Eqs.~(\ref{SL_P}) and~(\ref{SL_mu}), the corrections to these quantities, arising from the osmotic effect induced by the fillers, have been derived. These corrections are presented in Eqs.~(\ref{Pfil}) and~(\ref{mu1}), along with the expression for the chemical potential of the fillers given by Eq.~(\ref{mu3}). Based on these results, the spinodal conditions for both the unfilled [Eq.~(\ref{spin0})] and filled [Eq.~(\ref{spin})] polymer blends were derived. Under the assumption of low compressibility of the blend, the spinodal condition for the filled system simplifies to an analytical form given by Eq.~(\ref{spin_inc}). This approximate expression enables a direct analytical evaluation of the upper critical spinodal temperature $T(\phi)$ without the need for solving the full set of coupled nonlinear equations [Eqs.~(\ref{spin}) and~(\ref{Eq2_1})] that define the exact spinodal condition for a filled polymer blend.

Comparison between the exact spinodal and its incompressible-limit approximation (Fig.~\ref{fig3}) demonstrates that the low compressibility assumption introduces only minor deviations (less than $4~\mathrm{K}$), confirming the robustness of the developed approximation. Furthermore, fitting the theoretical spinodal to the experimental data (Fig.~\ref{fig4}) required only a small adjustment of the cross-interaction energy $\epsilon_{12}^0$, obtained from the Berthelot combining rule using the experimentally determined intra-species interaction energies $\epsilon_{ii}$ ($i = 1, 2$). The theoretical spinodals, both exact and approximate, show excellent agreement with experiment for filled and unfilled systems at small to moderate volume fractions $\phi$ of the lower–molecular-weight polymer (EVA). At larger $\phi$, the experimental data display a steeper decrease in $T(\phi)$ with increasing $\phi$ than that predicted theoretically. Nevertheless, the discrepancy between theoretical and experimental spinodal temperatures remains small, not exceeding $5~\mathrm{K}$ for the exact spinodal.

The derived analytical approximation for the spinodal of a filled polymer blend exhibiting UCST behavior provides a simple yet accurate means of estimating the spinodal temperature in polymer blends containing solid fillers. The developed approach can be readily extended to investigate the effects of filler geometry, polymer–filler interfacial interactions, and pressure dependence of the spinodal temperature on the miscibility behavior of polymer–particle mixtures. Such extensions are of considerable practical importance for the design of polymer nanocomposites with tailored phase stability, controlled through the optimized size, shape, and interfacial properties of the fillers.

In summary, the developed theory captures the essential thermodynamic features of weakly compressible polymer blends containing fillers and reproduces the experimental UCST spinodal data with high accuracy. The proposed analytical approximation provides a computationally efficient alternative to the full numerical evaluation of the spinodal in filled polymer blends and thus represents a valuable contribution to the predictive modeling of polymer–filler systems.

\appendix

\section{Derivation of the spinodal stability condition}
\label{app:A}
In this Appendix, we explicitly derive the derivatives $m_{ij} = M_j \partial_{M_j} \mu_1({M_i}, P)$ required for evaluating the stability condition given by Eq.~(\ref{spin}). It should be noted that these derivatives are taken at constant pressure $P$, which adds technical complexity to the calculation. Changing the independent variables from $(P,\{M_i\})$ to $(\eta,\phi,\varphi)$, with $P(\eta, \phi, \varphi) \equiv P_0 + \Delta P$ as given by Eqs.~(\ref{SL_P}) and~(\ref{Pfil}), one obtains, after straightforward but lengthy algebra, the following explicit expressions for $m_{ij}$.:
\begin{gather}
\label{m11}
m_{11} = r_1\left(
\phi_1 \phi_2^2 S_0 + 
	\phi_1 \left( 
	u_3 + \phi_2 \varphi (\tilde{\kappa}_0 \Delta\Pi + \Pi_{0}\Delta\tilde{\kappa} )
	\right) W_0 +
	\varphi \left( 
			u_1 \eta W_2 + \phi_1 \phi_2 W_3
			\right) 
\right) 
\end{gather}
\begin{gather}
\label{m13}
m_{13} = -r_1\left(
u_3 W_0 + \varphi (\eta u_3 W_2 - W_1)
\right) 
\end{gather}
\begin{gather}
\label{m31}
m_{31} = (u_1-\phi_1)(1 + \tilde{\kappa}_0/(1-u)) + \phi_1 W_1/(1-u)
\end{gather}
\begin{gather}
\label{m33}
m_{33} = 1- u_3(1 + \tilde{\kappa}_0/(1-u))
\end{gather}
Here, 
\begin{center}
\begin{gather}
\label{kappa}
\tilde{\kappa} = \tilde{\kappa}_0 + \varphi \Delta\tilde{\kappa} \\ \nonumber
\Delta\tilde{\kappa} =
1-u + 
2\eta \left(\frac{\phi}{r_1} + \frac{1 - \phi}{r_2} + \frac{(3 - 2\eta)\eta}{2(1 - \eta)^2} - 3\beta\eta\,\epsilon(\phi) \right)
\end{gather}
\end{center}
is the compressibility of {\it filled} polymer blend, $\tilde{\kappa}_0$ being the compressibility of pure polymer blend given by Eq.~(\ref{comp0}),
\begin{gather}
\label{PI}
\Pi\equiv \partial_{\phi} P = \Pi_{0} + \varphi \Delta \Pi , 
 \qquad 
\Delta \Pi = \eta^3 \left(r_1^{-1} - r_2^{-1} - 2 \beta \epsilon'(\phi) \right)
\end{gather}
is the derivative of the pressure with respect to the composition, and 
\begin{gather}
\label{W12}
W_0 = r_1^{-1} + \eta(1 - \eta)^{-1} - \beta \eta \left( 2\,\epsilon(\phi) + (1 - \phi)\,\epsilon'(\phi) \right) \\ \nonumber
W_1 = \tilde{\kappa}_0^{-1} + \phi_2\Pi, \qquad W_2 = \frac{\eta^2}{(1 - \eta)^2} - \frac{1}{r_1} + \frac{2 W_1}{\eta}, \\ \nonumber
W_3 = 
\eta \left( 1 + 2 \beta \eta \epsilon(\phi) - \beta \eta \phi_2^2 \epsilon''(\phi) - (1 - \eta)^{-1}\right) - r_1^{-1}\eta
\end{gather}
are  auxiliary functions. 

The obtained expressions for $m_{ij}$ are to be substituted to the spinodal condition given by Eq.~(\ref{spin}).  

\acknowledgments{Financial support of Deutsche Forschungsgemeinschaft (DFG) through Grant No. CH 845/2-3, is gratefully acknowledged.}


%

\end{document}